\documentclass[english]{iopart}

\begin{document}

\title[The diagonal Ising susceptibility]
{\Large
The diagonal Ising susceptibility
}

\author{ 
S. Boukraa$^\dag$, S. Hassani$^\S$, 
J.-M. Maillard$^\ddag$, B.M. McCoy$^*$ and N. Zenine$^\S$ }
\address{\dag LPTHIRM and D\'epartement d'A{\'e}ronautique,
 Universit\'e de Blida, Algeria}
\address{\S  Centre de Recherche Nucl\'eaire d'Alger, \\
2 Bd. Frantz Fanon, BP 399, 16000 Alger, Algeria}
\address{\ddag\ LPTMC, Universit\'e de Paris 6, Tour 24,
 4\`eme \'etage, case 121, \\
 4 Place Jussieu, 75252 Paris Cedex 05, France} 
\address{
$*$ C.N. Yang Institute for Theoretical Physics, \\
State University of New York, Stony Brook, NY, 11994, USA}
\ead{maillard@lptmc.jussieu.fr, maillard@lptl.jussieu.fr, mccoy@max2.physics.sunysb.edu,
boukraa@mail.univ-blida.dz, njzenine@yahoo.com}

\begin{abstract}
We use the recently derived form factor expansions of the diagonal two-point 
correlation function  of the square Ising model to study 
the susceptibility for a magnetic field applied only to one
diagonal of the lattice, for the isotropic  Ising model.
 We exactly evaluate the one and two particle
contributions $\chi_{d}^{(1)}$ and  $\chi_{d}^{(2)}$ of the 
corresponding susceptibility, and obtain linear differential 
equations for the three and four
particle contributions, as well as the five particle
 contribution ${\chi}^{(5)}_d(t)$, but only 
modulo a given prime. We use these exact linear differential equations 
to show that, not only the russian-doll
 structure, but also the direct sum 
structure on the linear differential operators
for the $\, n$-particle contributions $\chi_{d}^{(n)}$
 are quite directly inherited from the 
direct sum structure on the form factors  $\, f^{(n)}$. 
 We show that the $\, n^{th}$ particle contributions $\chi_{d}^{(n)}$
have their singularities at roots of unity. These singularities
 become dense on the unit circle $|\sinh2E_v/kT \sinh 2E_h/kT|=1$
as $ \, n\rightarrow \infty$.  
\end{abstract}

\vskip .3cm

\noindent {\bf PACS}: 02.30.Hq, 02.30.Gp, 02.30.-f, 
 02.40.Re, 05.50.+q, 05.10.-a, 04.20.Jb
\vskip .2cm
\noindent {\bf AMS Classification scheme numbers}: 
33E17, 33E05, 33Cxx,  33Dxx, 14Exx, 14Hxx, 34M55, 47E05,
 34Lxx, 34Mxx, 14Kxx
\vskip .1cm
 {\bf Keywords}: susceptibility of the isotropic square Ising model,  two-point 
correlation functions of the Ising model,  singularities
 of  the square Ising model,
 natural boundary, 
Fuchsian linear differential equations,  
 complete elliptic integrals.

\vskip .1cm

\section{Introduction}
\label{intro}
The magnetic susceptibility $\chi$ of the two dimensional Ising model
is expressed in terms of the two point correlation function
$C(M,N)=\, \langle\sigma_{0,0}\sigma_{M,N}\rangle$ as
\begin{eqnarray}
\label{sus}
kT \cdot
\chi \,= \,  \, \sum_{M=-\infty}^{\infty} \, \sum_{N=-\infty}^{\infty} 
(C(M,N)\, -{\cal  M}(0)^2)
\end{eqnarray}
where ${\cal M}(0)$ is the spontaneous magnetization (which vanishes
for $T> T_c$). This susceptibility has been studied for over three
decades by use of the form factor representations of the correlation
function~\cite{bm1}-\cite{nickel2}
\begin{eqnarray}
\label{cff}
C(M,N)_{\pm}\, =\,\, (1-t)^{1/4} \cdot \sum_{n}C^{(n)}(M,N)
\end{eqnarray}
where the subscript $+(-)$ denotes $T> T_c$ ($T< T_c$).
For $T>T_c$ the variable $t$ is
$t\,=\,\,(\sinh 2E_v/kT \sinh 2E_h/kT)^2$
and for $T<T_c$ is
$t\, =\,\, (\sinh 2E_v/kT \sinh 2E_h/kT)^{-2}$
where $E_v$ and $E_h$ are the vertical and horizontal interaction
constants. The sum is over
odd (even) values of the integer $n$ for  $T>T_c$ $(T<T_c)$
 and the $\, C^{(n)}(M,N)$ are explicit $n$ fold
integrals. Using this form factor decomposition (\ref{cff}) in the
expression for the susceptibility (\ref{sus}) we find
\begin{eqnarray}
\label{chisum}
kT \cdot \chi_{\pm}\, =\,\,  (1-t)^{1/4} \cdot \sum_{n}{\tilde \chi}^{(n)}
\end{eqnarray}
where the $ \, {\tilde \chi}^{(n)}$ can be expressed in terms of 
double sum of the form factors:  
\begin{eqnarray}
{\tilde \chi}^{(n)}\, =\,\,  \sum_{N=-\infty}^{\infty}\sum_{M=-\infty}^{\infty}
C^{(n)}(M,N) 
\end{eqnarray}

The study of these ${\tilde \chi}^{(n)}$ was initiated in \cite{bm1}
with the explicit evaluation of the integrals for  
${\tilde \chi}^{(1)}$ and ${\tilde \chi}^{(2)}$. However, 
a real understanding of the analytic structure
of ${\tilde \chi}^{(n)}$ began only in 1999 and 2000, with the
demonstration by Nickel \cite{nickel1, nickel2}, for the isotropic Ising model
($E_v=E_h=E$), that ${\tilde \chi}^{(n)}$ has a set of singularities,
lying on the unit circle $|\sinh 2E/kT|=1$, which become dense in
the limit $n\rightarrow \infty$. If these singularities do not cancel
in the full sum (\ref{chisum}) then the susceptibility will have a
{\em natural boundary} at $|\sinh2E/kT|=1$.

The existence of a natural boundary in the susceptibility is a
profound effect not envisioned in the traditional scaling, and
renormalization, description of critical phenomena. In order to obtain further
insight into  existence of such natural boundary,
 several of the present authors have made a
detailed study of ${\tilde \chi}^{(3)}$ and
 ${\tilde \chi}^{(4)}$ in~\cite{jm1}-\cite{jm4}.

The integrals for ${\tilde \chi}^{(n)}$, as explicitly 
written out in~\cite{nickel1, nickel2}, are quite complicated, and, for that
reason, it is difficult to extend the
analysis of~\cite{jm1}-\cite{jm4} to ${\tilde \chi}^{(n)}$ with $n\, \geq\, 5$.
A direct attack~\cite{jm5} on ${\tilde \chi}^{(5)}$ indicates that 
more than  6000 terms and, for ${\tilde \chi}^{(6)}$, probably more than
20000 terms in the power series expansion in $t$ are needed in order
to find the linear differential equations which they
satisfy. Therefore it would be most useful to study ``model
integrals'' that are simpler to analyze, and which incorporate
significant features of ${\tilde\chi}^{(n)}$.
Several such ``model integrals'' have been previously studied~\cite{jm6}
and revealed a rich structure of singularities beyond
 those found by Nickel~\cite{nickel1, nickel2}.
 
In this paper we introduce what is probably the most physical
simplification of the Ising susceptibility which retains the property
of having singularities on the unit circle $|t|=1$.
This model is obtained by considering the isotropic
 Ising model with a magnetic field
 {\em which acts only on one diagonal} of the lattice. 
The magnetic
susceptibility for this diagonal field will be, then, given by the diagonal
analogue of (\ref{sus}):
\begin{eqnarray}
\label{chid}
kT \cdot 
\chi_d \, =\,\, \,\, \sum_{N=-\infty}^{\infty} \, (C(N,N)\, -{\cal  M}(0)^2).
\end{eqnarray}
This expression\footnote[5]{Such partial sums have already been 
introduced and the asymptotic expansion of their short
term part considered in Orrick {\it et al.}~\cite{ongp}.} should
 be much easier to study than the
full susceptibility. This stems from
the form factor decomposition of the diagonal two-point correlations 
$\, C(N,N)$, that has been, recently,
presented~\cite{bm2} and proven~\cite{bm3}, and which is much simpler than
the decomposition obtained directly from~\cite{bm1}.
In particular for $T\, < \, T_c$ 
\begin{eqnarray}
\label{cdm}
C(N,N)_{-}\, =\, \,\, (1-t)^{1/4} \cdot
 \Bigl( 1+\sum_{n=1}^{\infty}\, f^{(2n)}(N,t) \Bigr)
\end{eqnarray}
with 
\begin{eqnarray}
\label{fdm}
&&f^{(2n)}(N,t) = \,\,\, \, \, { t^{n(n+|N|)} \over {
  (n!)^2 }} \, {{1 } \over {\pi^{2n} }} \,  
\int_0^1 \cdots \int_0^1\prod_{k=1}^{2n}\,  dx_k \cdot x_k^{ |N| }  
\nonumber \\ 
&&\quad \quad \quad \quad \times\prod_{j=1}^n\left({x_{2j-1}(1-x_{2j})(1-tx_{2j})
\over 
x_{2j}(1-x_{2j-1})(1\, -t\, x_{2j-1})}\right)^{1/2}\nonumber\\
&&\quad \quad \quad \quad \times \prod_{1 \leq j \leq n}
\prod_{1 \leq k \leq n}(1\, -t\, x_{2j-1}\, x_{2k})^{-2} \\
&&\quad \quad  \quad \quad \times
\prod_{1 \leq j<k\leq n}(x_{2j-1}-x_{2k-1})^2\, (x_{2j}-x_{2k})^2 \nonumber
\end{eqnarray}
and for $T>T_c$
\begin{eqnarray}
C(N,N)_{+}\, =\,\,  (1-t)^{1/4} \cdot \sum_{n=0}^{\infty}f^{(2n+1)}(N,t)
\label{cdp}
\end{eqnarray}
with
\begin{eqnarray}
\label{fdp}
&&f^{(2n+1)}(N,t)\,  = \,\,\,\,\,\,{t^{n(n+1)+ |N| (n+1/2)}
  \over \pi^{2n+1} n! (n+1)! } \cdot 
\int_0^1 \cdots  \, \int_0^1 \, \, \prod_{k=1}^{2n+1}\,  dx_k
 x_k^{ |N| }\nonumber\\
&&\quad\quad \quad \quad \times 
 \prod_{j=1}^{n}\, \Bigl((1-x_{2j})(1\,-t\,x_{2j})\cdot
 x_{2j}\Bigr)^{1/2} \nonumber \\
&&\quad \quad \quad \quad  \times\prod_{j=1}^{n+1} \, 
\Bigl((1\,  -x_{2j-1})(1\,-t\, x_{2j-1}) \cdot x_{2j-1}\Bigr)^{-1/2}
\nonumber \\
&&\quad \quad \quad \quad \times
\prod_{1\leq j\leq n+1}\prod_{1\leq k \leq n}\, 
(1\, -t\, x_{2j-1}\, x_{2k})^{-2} \\
&&\quad \quad  \quad \quad \times\prod_{1 \leq j<k\leq n+1}(x_{2j-1} -x_{2k-1})^2
\cdot \prod_{1\leq j<k\leq n}(x_{2j}-x_{2k})^2 \nonumber
\end{eqnarray} 

From now on, the integer $N$ should be understood as $\vert N \vert$
when evaluated.

 Thus, if we use (\ref{cdm}) and (\ref{cdp}) in (\ref{chid}), and
evaluate the sum on $N$ as a geometric series, we obtain for $ \, T < T_c$
\begin{eqnarray}
kT \cdot \chi_{d-}(t)\,\,  =\,\, \,  \, (1-t)^{1/4} \cdot 
\sum_{n=1}^{\infty} \, {\tilde \chi}_d^{(2n)}(t)
\label{defchidm}
\end{eqnarray}
with 
\begin{eqnarray}
&&{\tilde \chi}^{(2n)}_{d}(t)\,\, = \,\,\,  \, \,
 {{  t^{n^2}} \over {  (n!)^2 }} \, 
{{1 } \over {\pi^{2n} }} \cdot  
\int_0^1 \cdots  \,\int_0^1\prod_{k=1}^{2n}\,  dx_k  
\cdot   {1\, +t^n\, x_1\cdots x_{2n}\over
  1\,-t^n\, x_1 \cdots x_{2n}}\nonumber \\ 
&&\quad \quad \quad \quad \times
\prod_{j=1}^n\left({x_{2j-1}(1-x_{2j})(1-tx_{2j})\over 
x_{2j}(1-x_{2j-1})(1\, -t\, x_{2j-1})}\right)^{1/2}\nonumber\\
&&\quad \quad \quad \quad \times \prod_{1 \leq j \leq n}
\prod_{1 \leq k \leq n}(1\, -t\, x_{2j-1}\, x_{2k})^{-2} \nonumber\\
&&\quad\quad \quad \quad  \times
\prod_{1 \leq j<k\leq n}(x_{2j-1}-x_{2k-1})^2\, (x_{2j}-x_{2k})^2
\label{chidm}
\end{eqnarray}
and for $T>T_c$
\begin{eqnarray}
\label{defchidp}
kT \cdot \chi_{d+}(t)\, =\,\,\,  (1-t)^{1/4} \cdot
 \sum_{n=0}^{\infty}\, {\tilde \chi}_d^{(2n+1)}(t)
\end{eqnarray}
with 
\begin{eqnarray}
&&{\tilde \chi}^{(2n+1)}_{d}(t)\,\, = \,\,\,\,\,\,{t^{n(n+1))}
 \over \pi^{2n+1} n! \, (n+1)! } \cdot 
\int_0^1 \cdots \int_0^1 \, \, \prod_{k=1}^{2n+1}  dx_k\nonumber\\
&&\quad \quad \quad \quad  \times {1\, +t^{n+1/2}\,x_1\cdots x_{2n+1}\over
 1\, -t^{n+1/2}\, x_1\cdots x_{2n+1}} \cdot 
 \prod_{j=1}^{n}\, \Bigl((1-x_{2j})(1\,-t\,x_{2j})\cdot
 x_{2j}\Bigr)^{1/2} \nonumber \\
&&\quad \quad \quad \quad \times\prod_{j=1}^{n+1} \, 
\Bigl((1\,  -x_{2j-1})(1\,-t\, x_{2j-1}) \cdot x_{2j-1}\Bigr)^{-1/2} 
\nonumber\\
\label{chidp}
&&\quad \quad\quad  \quad  \times
\prod_{1\leq j\leq n+1}\prod_{1\leq k \leq n}\, 
(1\, -t\, x_{2j-1}\, x_{2k})^{-2} \\
&&\quad \quad  \quad \quad \times\prod_{1 \leq j<k\leq n+1}(x_{2j-1} -x_{2k-1})^2
\prod_{1\leq j<k\leq n}(x_{2j}-x_{2k})^2. \nonumber
\end{eqnarray} 
The expressions (\ref{chidm}) and (\ref{chidp}) are, indeed,
much simpler than the corresponding expressions for
${\tilde\chi}^{(n)}$ given in \cite{bm1}-\cite{jm4}.

In this paper we analyze the contributions ${\tilde \chi}^{(n)}_d(t)$
to the diagonal susceptibility. In section (\ref{evalu}) we analytically evaluate the
integrals for $n=1$ and $2$. In section (\ref{diffequ}) we present, and analyze, the
ordinary linear differential equations satisfied by ${\tilde \chi}^{(3)}_d(t)$
and ${\tilde \chi}^{(4)}_d(t)$, as well as  ${\tilde
  \chi}^{(5)}_d(t)$ modulo a given prime. These linear 
differential equations have a
direct sum decomposition which we relate in section (\ref{direct}) to the direct sum
decomposition found in~\cite{bm3} for the form factors
$f^{(n)}(N,t)$. In section (\ref{Singtilde}) we extract the singularities in the integral
representations (\ref{chidm}) and (\ref{chidp}) for ${\tilde \chi}^{(n)}_d(t)$. 
We find that all the singularities of 
${\tilde \chi}^{(2n)}_d(t)$  are at the roots of
unity $\, t^n\, =\, 1$, while, for ${\tilde \chi}^{(2n+1)}_d(t)$, they are at
$\, t^{n+1/2}\, =\, 1$.
These singularities for $t\neq 1$ are the counterparts, 
for ${\tilde \chi}^{(n)}_d(t)$, 
of the singularities on the unit circle found
 by Nickel~\cite{nickel1, nickel2} for the full susceptibility 
${\tilde \chi}$. However, unlike ${\tilde \chi}^{(n)}(t)$
which was shown in~\cite{jm1}-\cite{jm6} to have many other
singularities which lie outside the unit circle $|\sinh 2E/kT|=1$,
the diagonal ${\tilde \chi}^{(n)}_d(t)$ has no further singularities
other than those which satisfy $\, t^n\, =\, 1$ or $\, t^{n+1/2}\, =\, 1$.
We conclude in section (\ref{behavior}) 
with a discussion of 
the several different types of singularities in
$\, {\chi}_d(t)$ which follow from the singularities in  
${\tilde \chi}^{(n)}_d(t)$.

\section{Evaluation of  ${\tilde \chi}_d^{(1)}(t)$ and 
${\tilde \chi}_d^{(2)}(t)$}
\label{evalu}
The contribution of ${\tilde \chi}^{(1)}_d(t)$ is explicitly given
from (\ref{chidp}) as
\begin{eqnarray}
\label{chi10}
{\tilde \chi}^{(1)}_d(t)\, =\,\,\,\, {1\over \pi} \int_0^1 \, dx \cdot
  {1\, +t^{1/2}\,  x \over  1\, -t^{1/2}\, x}
 \cdot [(1-x)(1\, -t\, x)\,x]^{-1/2}
\end{eqnarray}
which, setting  $\, x\,=\,\, \sin^2\theta$ 
is written as
\begin{eqnarray}
\label{chi11}
{\tilde \chi}^{(1)}_d(t)\,=\,\,\,\,{2\over \pi}\int_0^{\pi/2}\, d\theta \cdot 
{1\, +t^{1/2}\sin^2\theta \over  1\, -t^{1/2}\sin^2\theta}
\cdot  (1\, -t\sin^2\theta)^{-1/2}.
\end{eqnarray}
Recalling the definitions of the elliptic integrals of the first kind
\begin{eqnarray}
K(t)\,=\,\,\, \int_0^{\pi/2}{d\theta \over (1\,-t \,\sin^2\theta)^{1/2}}
\end{eqnarray}
and of the third kind
\begin{eqnarray}
\Pi_1(\nu,t)\,=\,\,\,\, \int_0^{\pi/2}
{d\theta \over (1\,+\nu  \sin^2\theta)(1\,-t\, \sin^2\theta)^{1/2}}
\end{eqnarray}
we may write (\ref{chi11}) as
\begin{eqnarray}
{\tilde \chi}^{(1)}_d(t)\, =\, \, \,\, \, {2 \over  \pi}\cdot 
 \left(2\, \Pi_1(-t^{1/2},\,t) \,\,\, -K(t)\right).
\end{eqnarray}
Thus, if we use the identity \cite{cayley}
\begin{eqnarray}
\label{identPi1}
&&{{2} \over {\pi}} \cdot \Bigl(\Pi_1(\nu,\,t)\,+\Pi_1(t/\nu,\, t) \Bigr) 
\,\, =\,\, \\
&& \qquad \qquad  \qquad {{2} \over {\pi}} \cdot  K(t)\, \,  
+ \, [(1+\nu)(1+t/\nu)]^{-1/2} \nonumber 
\end{eqnarray}
with $\,\nu=\,-t^{1/2}$ we find:
\begin{eqnarray}
\label{chi12}
{\tilde \chi}^{(1)}_d(t)\,=\,\,\,\, {1 \over 1\,\, -t^{1/2}}.
\end{eqnarray}

It is instructive to derive (\ref{chi12}) without recourse to
identities on elliptic integrals. We first 
rewrite (\ref{chi10}) as a contour integral
\begin{eqnarray}
\label{chi14}
{\tilde \chi}_d^{(1)}(t)\,=\,\,\,\,
 {1 \over 2\pi i}\,  \oint{dz \over z} \cdot {1\,+z\over 1-z}
\cdot
[(1-t^{1/2}z)(1-t^{1/2}z^{-1})]^{-1/2}
\end{eqnarray}
on the contour $|z|<1$ (which is, in fact, the form in which the form
factors $\, f^{(n)}(N,t)$ were originally derived~\cite{bm3}). The
integrand in (\ref{chi14}) is antisymmetric if we send
 $ z \rightarrow1/z$. Therefore
\begin{eqnarray}
\label{chi15}
{\tilde \chi}^{(1)}_d(t)\, =\, \, \,\, \,-{\tilde \chi}^{(1)}_d(t)\,\,\,
 -({\rm  residue~~at~~}z=1)
\end{eqnarray}
The residue at $z=1$ is easily evaluated as
\begin{eqnarray}
\label{chi16}
({\rm  residue~~at~~}z=1)\,\, =\, \,\, \,\, -{2 \over 1-t^{1/2}}
\end{eqnarray}
and thus using (\ref{chi16}) in (\ref{chi15}) we again obtain the
result (\ref{chi12}).

To compute ${\tilde \chi}^{(2)}_d(t)$ we also use the contour integral
method, and rewrite (\ref{chidm})  with $n=1$ as
\begin{eqnarray}
\label{chi21}
&& {\tilde \chi}^{(2)}_d(t)\, =\,\, \,\,
{1\over (2 \pi i)^2}\oint dz_1 \oint dz_2 \\
&& \qquad \qquad \times {1\, +z_1 z_2 \over (1\, -z_1 z_2)^3}
 \cdot \left[{(1\, -t^{1/2}z_2)(1\, -t^{1/2}z_2^{-1})\over
    (1\, -t^{1/2}z_1)(1\, -t^{1/2} z_1^{-1})}\right]^{1/2}
 \nonumber 
\end{eqnarray}
on the contour $|z_k|<1$ for $k=1,2$.
As was the case for ${\tilde \chi}^{(1)}_d(t)$, we note that the
integrand of (\ref{chi21}) is antisymmetric if we send $z_1\rightarrow
1/z_1,~z_2\rightarrow 1/z_2$. Therefore
\begin{eqnarray}
\label{chi22}
{\tilde \chi}^{(2)}_d(t)\, =\,\,\, 
-{\pi i\over (2\pi i)^2}\oint dz_1({\rm  residue~~at~~}z_2=z_1^{-1}) 
\end{eqnarray}
where:
\begin{eqnarray}
\label{chi23}
&&({\rm residue~~at~~}z_2=\, z_1^{-1}) \,
=\, \,\, z_1^{-3}[(1-t^{1/2}z_1)(1\, -t^{1/2} z_1^{-1}]^{-1/2} \\
&&\qquad\,  \quad \times{1 \over 2}{d^2\over
  dz_2^2}\{(1+z_1z_2)[(1\, -t^{1/2}z_2)
(1\, -t^{1/2}z_2^{-1}]^{1/2}\}|_{z_2= z_1^{-1}}.  \nonumber
\end{eqnarray}
When (\ref{chi23}) is evaluated, and substituted into (\ref{chi22}), the
resulting integral over $z_1$ has only poles. Keeping only those
terms which give nonvanishing contributions, we find
\begin{eqnarray}
\label{chi24}
{\tilde \chi}^{(2)}_d(t)
\,\,=\,\,\, {1\over 8\pi i}\oint dz_1 
\,{t \over (1\,-t^{1/2}z_1)(z_1\,-t^{1/2})}
\,\,=\,\,\,\,{t \over 4\,(1-t)}. 
\end{eqnarray}

It should be noted that neither ${\tilde \chi}^{(1)}_d(t)$,
nor ${\tilde \chi}^{(2)}_d(t)$, contain logarithms even though there
are logarithms in both $f^{(1)}(N,t)$ and $f^{(2)}(N,t)$ for all
$N$. This is to be contrasted with the corresponding results for the
full susceptibility where it was seen in~\cite{bm1} that 
${\tilde \chi}^{(n)}(t)$ has no logarithms for $n=1$, but does have a term in
$\ln t$ for $n=2$.

\section{Linear differential equations for ${\tilde \chi}_d^{(3)}(t)$,
${\tilde \chi}_d^{(4)}(t)$  and ${\tilde \chi}_d^{(5)}(t)$}
\label{diffequ}
We now turn to ${\tilde \chi}_d^{(n)}(t)$ for $n\geq 3$. When written
in contour integral form the integrands still have the property of
being antisymmetric when $z_k\, \rightarrow\, 1/z_k$. Unfortunately this
property is not sufficient to reduce the computation to integrals
that can all be evaluated by residues. Therefore we do not have an
explicit evaluation in terms of elementary functions, and we continue
our study by using formal computer computations to determine the
linear differential equations satisfied by ${\tilde
  \chi}_d^{(3)}{(t)}$, ${\tilde \chi}_d^{(4)}(t)$ and ${\tilde \chi}_d^{(5)}(t)$, as was done for
${\tilde \chi}^{(3)}$ and ${\tilde \chi}^{(4)}$ for the full
susceptibility in \cite{jm1}-\cite{jm4} and for the $f^{(n)}(N,t)$ in
\cite{bm2}. We present the results of these computer calculations for 
${\tilde \chi}_d^{(3)}{(t)}$,
${\tilde \chi}_d^{(4)}(t)$ and ${\tilde \chi}_d^{(5)}(t)$ separately.

\subsection{Linear differential equation for ${\tilde \chi}_d^{(3)}(t)$}
\label{diffequtilde3}
For $\, \chi^{(3)}_d$ we chose $x= \,t^{1/2}\, =\, \sinh 2E_v/kT \, \sinh 2E_h/kT$ as
  our independent variable. We find that the linear differential operator for
  ${\tilde \chi}_d^{(3)}(x)$ is of order six, and has   
the direct sum decomposition
\begin{eqnarray}
{\cal L}_6^{(3)}\,= \,\,\, \, L_1^{(3)} \oplus L_2^{(3)} \oplus L_3^{(3)}
\label{ds3}
\end{eqnarray}
with
\begin{eqnarray}
L_1^{(3)}\, =\,\,\,\, Dx\, + \,\, {\frac{1}{x-1}}, 
\end{eqnarray}
\begin{eqnarray}
L_2^{(3)} =\,\,\,
Dx^{2}\,\, +2\,{\frac { (1+2\,x)}{ \left( 1+x
 \right)  \left( x-1 \right) }} \cdot Dx
\,+{\frac {1+2\,x}{(1+x)  \, (x-1)\, x }},
\end{eqnarray}
\begin{eqnarray}
&& L_3^{(3)} =\,\,\,\, Dx^{3}\, \\ 
&&\qquad \quad +  {{3} \over {2}}\,{\frac { \left( 8\,{x}^{6}+36\,{x}^{5}+63\,{x}^{4}
+62\,{x}^{3}+21\,{x}^{2}-6\,x-4 \right)}{ \left( x+2
 \right)  \left( 1+2\,x \right)  \left( 1+x \right)  
\left(x-1 \right)  \, (1+x+x^2)\, x}} \cdot Dx^{2} \nonumber \\
&&\qquad \quad  +{\frac { n_1 }{ \left( x+2 \right)  \left(1+2\,x \right)
  \left( 1+x \right) ^{2} \left( x-1 \right)^{2} 
\left(1+x+x^2 \right) {x}^{2}}} \cdot  Dx \nonumber \\
&&\qquad \quad  +{\frac { n_0}{ (x+2)  \,(1+2\,x)  \,(x-1)^{3} \, (1+x+x^2) 
 \, (1+x)^{2}{x}^{2}}} \nonumber
\end{eqnarray}
with:
\begin{eqnarray}
&&n_0 \, = \, \,
2\,{x}^{8}+8\,{x}^{7}-7\,{x}^{6}-13\,{x}^{5}
-58\,{x}^{4}-88\,{x}^{3}-52\,{x}^{2}-13\,x+5,   \nonumber \\
&&n_1 \, = \, \, 
14\,{x}^{8}+71\,{x}^{7}+146\,{x}^{6}+170\,{x}^{5}
 +38\,{x}^{4}  \nonumber \\
&&\quad \quad \quad \quad \quad -112\,{x}^{3}-94\,{x}^{2}-19\,x+2.   \nonumber
\end{eqnarray}

The solution of $\, L_1^{(3)}$ is in fact (up to a constant)
${\tilde \chi}_d^{(1)}(t)$ as given in (\ref{chi12}).
This means that  ${\tilde \chi}_d^{(1)}(t)$ is actually solution 
of the Fuchsian linear differential operator 
 $\, {\cal L}_6^{(3)}$ corresponding to  ${\tilde \chi}_d^{(3)}(t)$.
We recover then the ``russian-doll structure'' noticed in~\cite{jm3,jm4}
for the third contribution  $\, {\tilde \chi}^{(3)}$
to the full susceptibility.

The linear differential operator of order two, $L_2^{(3)}$ has, at $x=0,1,-1$
and $x=\infty$, respectively  the exponents $(0,1),(-2,0), (0,0),$ and
$(1,2)$. Some manipulations on the formal solutions, at $x=\, 0$, give 
the result that the solution analytic at $x=0$ reads
\begin{eqnarray}
\label{solL2}
sol(L_2^{(3)})\,  =\, \, \,  {\frac{1}{x-1}}\cdot  K(x^2)\,\, 
 + \, {\frac{1}{(x-1)^2}}\cdot  E(x^2)
\end{eqnarray}
with:
\begin{eqnarray}
\label{KE}
K(y) =\, F \left( 1/2, 1/2; 1; y \right), \quad \, \,  \, \,
E(y) =\, F \left( 1/2, -1/2; 1; y \right)
\end{eqnarray}
(where we have abused conventional notation by omitting the factor of $\pi/2$
in the canonical definition\footnote[2]{In maple's notations, for
 $ t \, = \, k^2$ ($k$ is the modulus), 
$\, K(t) \, = \, K(k^2)$ in (\ref{KE}) 
 reads : $\, hypergeom( [1/2,1/2],[1],t)$  
 $\,= \,  2/\pi \cdot EllipticK(k)$, but reads $ 2/\pi \cdot EllipticK[k^2]$ 
in Mathematica.} of the complete elliptic integrals
 K and E).

We have not found the explicit solution of the linear
 differential operator $\, L_3^{(3)}$
which has the following regular singular points and exponents:
\begin{eqnarray}
\label{l33exp}
1+x+x^2 = 0,&\quad\quad \rho = 0, 1, 7/2 & \rightarrow 
 \quad \quad \quad \,\,x^{7/2},  \nonumber \\
x = ~~ 0&\quad \quad \rho = 0, 0, 0 & \rightarrow 
\quad \quad\quad \,\, log^2\,\,\,\, terms, \nonumber \\
x = ~~1 &\quad \quad  \rho = -2, -1, 1 & \rightarrow 
\quad \quad \quad \,\, x^{-2}, \,\, x^{-1}, \\ 
x = -1&\quad \quad \rho = 0, 0, 0 & \rightarrow 
\quad \quad \quad \,\,log^2\,\,\,\, terms,  \nonumber \\
x = ~\infty& \quad \quad \rho = 1, 1, 1 & \rightarrow 
\quad \quad \quad \,\, log^2\,\, \,\,terms.  \nonumber
\end{eqnarray}
The singularities at $x=\, 2,\, -1/2$ are apparent.

\subsection{Linear differential equation for ${\tilde \chi}_d^{(4)}(t)$} 
\label{diffequtilde4}
For ${\tilde \chi}_d^{(4)}(t)$ we use $t$ as the independent variable.
The  linear differential operator for ${\tilde \chi}^{(4)}_d(t)$ 
is of order eight, and has the direct sum decomposition
\begin{eqnarray}
{\cal L}_8^{(4)}\,=\,\,\, \,  L_1^{(4)} \oplus L_3^{(4)} \oplus L_4^{(4)}
\label{ds4}
\end{eqnarray}
with
\begin{eqnarray}
L_1^{(4)}\,=\,\,\, Dt\, +{\frac{1}{(t-1)\, t}},
\end{eqnarray}
\begin{eqnarray}
&&L_3^{(4)}\, = \,\,   Dt^{3}\, 
+{\frac { \left( 5\,{t}^{2}+6\,t-1 \right) }
{ \left( 1+t \right)  \, (t-1)\, t }} \cdot Dt^{2}
\,\, +{\frac { \left( 3\,{t}^{3}+6\,{t}^{2}-2\,t-1 \right)}
{ \left( 1+t \right) {t}^{2}
 \left( t-1 \right) ^{2}}}\cdot  Dt \nonumber\\
&&\qquad \qquad \qquad -{\frac {3}{ 2\, (1+t) \, (t-1)\, {t}^{2} }}, 
\end{eqnarray}
\begin{eqnarray}
&&L_4^{(4)}\,  =\, \,  Dt^{4}\, +{\frac { \left( 7\,{t}^{4}
-68\,{t}^{3}-114\,{t}^{2}+52\,t-5 \right) }{ (t+1) 
 \, ({t}^{2}-10\,t +1) \, (t-1)\,  t  }}  \cdot Dt^{3} \nonumber\\
&&\qquad \qquad \,  + 2\,{\frac { \left( 5\,{t}^{5}
-55\,{t}^{4}-169\,{t}^{3}+149\,{t}^{2}-28\,t+2 \right)}{
 \left( t+1 \right)  \left( {t}^{2}-10\,t+1 \right) 
{t}^{2} \, (t-1)^{2}}} \cdot  Dt^{2} \nonumber \\
&&\qquad \qquad \, + 2\,{\frac { \left( {t}^{4}
-13\,{t}^{3}-129\,{t}^{2}+49\,t-4 \right)}
{ (t+1)  \left( {t}^{2}-10\,t+1\right) {t}^{2}
 \left( t-1 \right)^{2}}} \cdot  Dt  \\
&&\qquad \qquad \, -3\,{\frac { \left( t+1 \right)^{2}}
{ \left( {t}^{2}-10\,t+1 \right)  \left( t-1 \right)^{2 }{t}^{3}}}. \nonumber
\end{eqnarray}

The solution of $\,L_1^{(4)}$ is (up to a constant) the function 
${\tilde \chi}_d^{(2)}(t)$ given in (\ref{chi24}).
Here also, ${\tilde \chi}_d^{(2)}(t)$ is solution of 
the Fuchsian linear differential operator 
 $\, {\cal L}_8^{(4)}$ corresponding to  ${\tilde \chi}_d^{(4)}(t)$,
and we recover, again, the russian-doll structure noticed
in~\cite{jm3,jm4} for the fourth contribution ${\tilde \chi}^{(4)}$
to the full susceptibility.

The solution of $\, L_3^{(4)}$ is found to be (with notations (\ref{KE})):
\begin{eqnarray}
sol(L_3^{(4)})\,  = \, \, \,  -K(t)^2\,\,   
  + {\frac{1+t}{(1-t)^2}}\cdot  E(t)^2\,\,  
 +\,   {\frac{2t}{t-1}}\cdot  K(t)\,E(t).
\end{eqnarray}

We have not found the explicit solution for the operator $\, L_4^{(4)}$
which has the following regular singular points and
exponents
\begin{eqnarray}
\label{l44exp}
t = ~~0,& \quad \quad \quad \rho = 0, 0, 0, 1 & \quad \quad   \rightarrow \quad 
\quad \quad  log^3\,\, \,\, \, \,  terms, \nonumber \\
t = ~~1,& \quad\quad \quad  \rho = -2, -1, 0, 1 &
\quad \quad \rightarrow \quad \quad \quad 
 t^{-2},\, \, t^{-1},\, \, log\,\,\,\, \,  \, term, \nonumber \\
t = -1,& \quad \quad\quad  \rho = 0, 1, 2, 7 & \quad \quad \rightarrow 
\quad \quad \quad   t^7\, log\,\,\, \,  \, term, \\
t = ~\infty, & \quad \quad \quad \rho = 0, 0, 0, 1 &\quad \quad \rightarrow 
\quad \quad  \quad  log^3 \,\,\,\, \,  \, terms. \nonumber
\end{eqnarray}
The singularities at the roots of $\,\, t^2\, -10t\, +1 =\,0\,$ are 
apparent.

\subsection{Linear differential equation for ${\tilde \chi}_d^{(5)}(t)$}
\label{Diffeqtilde5}
For $\, \tilde{\chi}^{(5)}_d$ we chose, again,
 $x=\, t^{1/2}\, =\, \sinh2E_v/kT \, \sinh 2E_h/kT$ as
  our independent variable.
The first terms of the series expansion of $\, {\tilde \chi}_d^{(5)}(x)$ read :
\begin{eqnarray}
\label{first}
&& {\tilde \chi}_d^{(5)}(x) \, = \, \,\, {{3} \over {262144}} \cdot x^{12} 
\, \,+ \,{{39} \over {1048576}} \cdot x^{14}
 \,\, + \,{{5085} \over {67108864}} \cdot x^{16} \, \\
&& \quad \quad \quad  + \, {{9} \over {67108864}} \cdot x^{17}\,
+ \,{{33405} \over {268435456}} \cdot x^{18} \,
 + \,{{ 315} \over {536870912}} \cdot x^{19} \, + \cdots \nonumber
\end{eqnarray}
Now, the calculations, in order to get the linear differential operator for
 $\, {\tilde \chi}_d^{(5)}(x)$, become really huge.
 For that reason, we introduce a ``modular'' strategy
which amounts to generating large series  {\em modulo  
a given prime}, and then deduce, 
from a Pad\'e-Hermite procedure, the linear differential operator  for
  $\, {\tilde \chi}_d^{(5)}(x)$ {\em modulo that prime}. 
We have generated  $\,3000 $ coefficients for the 
series expansion of $\, {\tilde \chi}_d^{(5)}(x)$ modulo a given prime
(here $32003$), and actually found
a linear differential equation {\em modulo that prime}
of order $\, 25,\, 26, \, \cdots$,  using $\,2200 $
 terms in the series expansion.
One can also obtain linear differential equations of smaller
 order for  $\, {\tilde \chi}_d^{(5)}(x)$, but where
{\em more terms} ($2500,\, 2600, \, 2800, \cdots$), are needed.
The polynomial corresponding to apparent singularities in front
of the highest derivative is now very large.
The smallest order one can reach is $\, 19$, and the
linear differential equation we have ``guessed'' has required
more than $\,3000 $ terms in the series expansion.
Note that a linear differential equation of the minimal order
is not, necessarily, the simplest one,
as far as the number of terms in the series expansion needed to
guess it, is concerned.
We have already encountered such a situation in~\cite{jm4}. 

Recalling $\, {\cal L}_6^{(3)}$, the  order 
six linear differential operator corresponding to 
${\tilde \chi}_d^{(3)}(x)$, one finally finds that the
linear differential operators $\,  {\cal L}_n^{(5)}\,$
 we have obtained\footnote[2]{The 
method consists in searching, modulo a given prime, the linear differential operator
for $\,{\cal L}_6^{(3)}({\tilde \chi}_d^{(5)})$.} for  ${\tilde \chi}_d^{(5)}(x)$,
have the following factorization:
\begin{eqnarray}
\label{russian}
{\cal L}_n^{(5)} \, \,  \,= \, \, \, \, \, 
{\cal L}_{n-6}^{(5)} \cdot {\cal L}_6^{(3)}
\end{eqnarray}
implying that  ${\tilde \chi}_d^{(3)}(x)$ is actually a 
solution of  $\,  {\cal L}_n^{(5)}$,
the linear differential operator
for  ${\tilde \chi}_d^{(5)}(x)$. 
The ``russian-doll'' structure shows up again.
For the linear differential operator $\, {\cal L}_{n}^{(5)} $
of smallest order ($n=19$),  $\, {\cal L}_{n-6}^{(5)} $
has a polynomial of apparent singularities of degree $\, 331$.
For larger orders for $\, {\cal L}_{n-6}^{(5)} $ we 
get smaller  apparent polynomials.
In \ref{B}, we sketch an order twenty
linear differential operator $\, {\cal L}_{n-6}^{(5)} $ 
modulo the prime $\, 32003$, which has 
{\em no apparent singularities}, and only  
the ``true'' singularities of 
the linear differential operator $\,  {\cal L}_n^{(5)}\,$
for  ${\tilde \chi}_d^{(5)}(x)$:
\begin{eqnarray}
\label{truesing}
(x+1) \cdot (x-1)  \cdot(x^2+x+1) \cdot (x^4+x^3+x^2+x+1) \
\cdot x  \, = \,\, \, 0. \nonumber
\end{eqnarray}
We have not yet been able to get the direct sum 
decomposition of $\, {\cal L}_n^{(5)}$ in this approach modulo prime.

\section{Singularities of ${\tilde \chi}_d^{(n)}(t)$} 
 \label{Singtilde}

In~\cite{bm2} we found that the form factors $f^{(n)}(N,t)$ 
have singularities of the form $\ln^n(1-t)$ at $t\rightarrow 1$. These
singularities come from the factors $[(1-x_j)(1-tx_j)]^{1/2}$ in
the integrands of (\ref{fdm}) and (\ref{fdp}). These factors are also
present in the integrands (\ref{chidm}) 
and (\ref{chidp}) for ${\tilde \chi}_d^{(n)}(t)$, and, thus, we 
expect that, for general values of $n$,
there will be powers of $\ln(1-t)$ present in ${\tilde
  \chi}_d^{(n)}(t)$.
We see, from the previous section,  that these logarithmic 
singularities are first seen in the linear differential
equation for ${\tilde \chi}_d^{(3)}(t)$.

However, there are additional singularities in 
 ${\tilde \chi}_d^{(2n)}(t)$ coming from the vanishing of the denominator 
\begin{eqnarray}
1\,\, - \, t^n \, x_1\, x_2 \, \cdots\, x_{2n},
\label{unit1}
\end{eqnarray}
and, in ${\tilde \chi}_d^{(2n+1)}(t)$, coming from the vanishing of the denominator 
\begin{eqnarray}
\label{unit2}
1\, \,-\,t^{n+1/2}\, x_1 \, x_2\, \cdots \, x_{2n+1}
\end{eqnarray}
which are not present in $\, f^{(n)}(t)$. For ${\tilde \chi}_d^{(2n)}$
the vanishing of (\ref{unit1}) occurs for  $ \, t^n=1 $
at the endpoints $\, x_1=\,x_2=\,\cdots\, =\, x_{2n}=1$,
and, for ${\tilde \chi}_d^{(2n+1)}$,
the vanishing of (\ref{unit2}) occurs for $ \, t^{n+1/2}=1$ 
at the endpoints  $\, x_1=\, x_2=\, \cdots\,  =\, x_{2n+1}=1$.  
 
When $\,t\, \rightarrow\, 1$ this additional singularity is the simple pole 
$(1-t)^{-1}$ for both ${\tilde \chi}_d^{(2n)}(t)$ and 
${\tilde \chi}_d^{(2n+1)}(t)$. When $\, t$ 
approaches the roots of unity $ t^n=1$ (with $ \, t \neq\,  1$) 
\begin{eqnarray}
\label{unit7}
t_{l,n}\, =\,\,\, e^{2\pi il/n} 
\quad \quad {\rm with} \quad \quad l \, =\, 1,\, 2,\, \cdots,\, n-1,
\end{eqnarray} 
then $\, {\tilde \chi}_d^{(2n)}(t)$ has a singularity of the form
\begin{eqnarray}
\label{unit8}
\kappa_{2n} \cdot (t-t_{l,n})^{2n^2-1} \cdot \ln (t-t_{l,n}).
\end{eqnarray}
Similarly, when $t$ approaches the roots of unity 
for  ${\tilde \chi}_d^{(2n+1)}$ (with $t\neq 1$), namely $\, t_0^{n+1/2}=1$, 
then ${\tilde \chi}_d^{(2n+1)}(t)$ has a singularity 
of the form  $\, \kappa_{2n+1} \cdot (t^{1/2}-t_0^{1/2})^{(n+1)^2-1/2}$.

These singularities, on the unit circle in the complex $t$ plane (for
$t \,\neq \, 1$), are the counterparts for the diagonal susceptibility
of the singularities
found by Nickel~\cite{nickel1, nickel2} on the unit circle
$|s|=1$, with $s=\, \sinh 2E/kT$, for the ${\tilde \chi}^{(n)}$ of 
the isotropic Ising model.

There are {\em no other} values of $t$ in the complex plane for
which the functions ${\tilde \chi}_d^{(n)}(t)$ are singular. This is
in distinct contrast with the ${\tilde \chi}^{(n)}(s)$ of the
isotropic Ising model which have singularities at many other places on
the complex $s$ plane~\cite{jm1}-\cite{jm4},\cite{jm6}. In this
section we derive the behavior of ${\tilde \chi}^{(n)}_d(t)$ at the
root of unity points  $ \, t^{n}=1$ and  $ \, t^{n+1/2}=1$.
Appendix B gives the values of the amplitudes at the singular points
lying on the unit circle $\vert t \vert=1$, 
for ${\tilde \chi}_d^{(3)}$ and  ${\tilde \chi}_d^{(4)}$, using
the matrix connection method \cite{jm4}.

\subsection{The singularity in ${\tilde \chi}_d^{(n)}$ at $t=1$}
\label{singunt1}

To extract the dominant singularity in ${\tilde \chi}_d^{(2n)}$ at $t=\, 1$
we set $\, t\, =\,\, 1-\epsilon\, $ and $\, x_k\, =\, \, 1\, -\epsilon \cdot  y_k$
 in (\ref{chidm}), and set $\epsilon=\, 0$ wherever possible. Thus we obtain 
the result that as $t\rightarrow 1$ ($\epsilon \rightarrow 0$)
\begin{eqnarray}
\label{unit13a}
{\tilde \chi}_d^{(2n)}\, \,\, \sim\,\,\, {{1} \over {1-t}}  \cdot I^{(2n)}_d 
\end{eqnarray}
with: 
\begin{eqnarray}
\label{unit13}
&&I_d^{(2n)}\, =\, 
{2 \over (n!)^2\, (2\pi)^{2n}}
\int_{0}^{\infty} dy_1 \cdots \int_0^{\infty}dy_{2n}
\cdot \prod_{j=1}^{n}\left({y_{2j}(1+y_{2j})\over
  y_{2j-1} \cdot (1+y_{2j-1})}\right)^{1/2}\nonumber\\
&&~~~~~~~~~~~\quad \times {1\over n+y_1+\cdots +y_{2n}}
\prod_{1\leq j\leq n}\prod_{1\leq k\leq n}(1+y_{2j-1}+y_{2k})^{-2}\nonumber\\
&&~~~~~~~~~~~\quad\times \prod_{1\leq j<k\leq n}\, 
(y_{2j-1} -y_{2k-1})^2(y_{2j}-y_{2k})^2.
\end{eqnarray}

Similarly, we find the singularity in ${\tilde\chi}_d^{(2n+1)}$ at $t=1$
reads
\begin{eqnarray}
\label{unit14a}
{\tilde \chi}_d^{(2n+1)} \, \, \sim \,  \, \, {{1} \over {1-t}}  \cdot I^{(2n+1)}_d 
\end{eqnarray}
with:
\begin{eqnarray}
\label{unit14}
&&I^{(2n+1)}_d\,  =\, \, \, {2 \over n! \, (n+1)! \,(2\pi)^{2n+1}}
\int_{0}^{\infty}\,  dy_1 \cdots \int_0^{\infty} dy_{2n+1}
\nonumber\\
&& \qquad  {1 \over n+y_1+\cdots +y_{2n}} \, 
\prod_{j=1}^{n}[(1+y_{2j}) \, y_{2j}]^{1/2}\,
 \prod_{j=1}^{n+1}[ y_{2j-1}(1+y_{2j-1})]^{-1/2}
\nonumber\\
&&\qquad \times \prod_{1\leq j\leq n+1} 
\prod_{1\leq k\leq n}(1 +y_{2j-1} +y_{2k})^{-2}\nonumber\\
&&\qquad \times \prod_{1\leq j<k\leq n+1}(y_{2j-1}-y_{2k-1})^2 \cdot 
\prod_{1\leq j<k\leq n}(y_{2j}-y_{2k})^2. 
\end{eqnarray}

\subsection{Singularities for ${\tilde \chi}_d^{(2n)}(t)$ 
at $t^n=1$  for $\, t\, \neq\, 1$} 
\label{singchi2n}
When $n=2$ the root of unity singularities  $ \, t^{n}=1$,
which is not $t=1$, occurs at
$t=-1$ where the analysis of the differential equation given 
in section (\ref{diffequ})
shows that there is a singularity of the form $\, (1+t)^7 \cdot \ln (1+t)$,
(see Appendix B).

To demonstrate, for general $n$, that the singularity in 
${\tilde  \chi}_d^{(2n)}(t)$, at the points $ \, t^{n}=1$, is given by 
(\ref{unit8}),
we set $\,\, t\,=\,\, t_{l,n}\cdot (1-\epsilon)$ and $\,x_k\,=\,1\,-\epsilon \cdot  y_k$.
Furthermore, because the singularity only occurs in a high derivative of
${\tilde \chi}^{(2n)}_d(t)$, we consider the $m^{th}$ derivative of 
${\tilde \chi}^{(2n)}_d(t)$, and will eventually see that $\,m$ should be
chosen to be $\,2\,n^2\,-1$. Then using
 $\,t\,=\,\, t_{l,n}\cdot (1-\epsilon)$ and $\,x_k\,=\, 1\,-\epsilon \cdot  y_k$, in
the $m^{th}$ derivative of (\ref{chidm}), and setting $\, t=t_{l,n}$ and
$x_k\, =\,1$ wherever possible, we obtain a model integral whose leading
singularity, at $t=\, t_{l,m}$, will be the same as the leading singularity
in the $\, m^{th}$ derivative of ${\tilde \chi}_d^{(2n)}(t)$:
\begin{eqnarray}
&&I^{(2n)}_m(t)\,\,=\,\,\,\, {\epsilon^{2n^2-1-m}\, 2m!\over
      (n!)^2 \,\pi^{2n} \, (1-t_{l,n})^{n^2}} \cdot 
\int_0^{\infty} \cdots
      \int_0^{\infty}\prod_{j=1}^n dy_j \\
 && \qquad \qquad \times ({y_{2j-1}\over
      y_{2j}})^{1/2} \cdot {1\over (n+y_1+y_2+\cdots +y_{2n})^{m+1}} 
\nonumber \\
 && \qquad \qquad \times \prod_{1\leq j<k\leq n}
(y_{2j-1}-y_{2k-1})^2(y_{2j}-y_{2k})^2. \nonumber
\end{eqnarray}
This expression is formally independent of $\epsilon$ when $m=\, 2n^2-1$.
When $\, m$ has this value the integral diverges logarithmically when
all $y_k$ become large, such that all ratios $y_j/y_k$ are of order
one. We thus conclude that the singularity in 
${\tilde\chi}^{(2n)}_d(t)$, at $t=\, t_{l,n}$, is given by (\ref{unit8}). 

\subsection{Singularities for ${\tilde \chi}_d^{(2n+1)}(t)$ at
  $\, t^{n+1/2}=1$ for $\, t\neq 1$} 
\label{Singulnplus1/2}
For the function ${\tilde \chi}_d^{(3)}(t)$ we used in section
 (\ref{diffequ}) the
independent variable $x=\, t^{1/2}$, and found, at the points $x_0^2\, +x_0\, +1=\, 0$, 
that there is a singularity of the form  $\, (x-x_0)^{7/2} $, (see Appendix B).
An analysis, completely analogous to the analysis given above for
${\tilde \chi}^{(2n)}(t)$, demonstrates that, for all $n$, 
the singularity in ${\tilde \chi}_d^{(2n+1)}(t)$, at 
$\, x_0^{2n+1}=t_0^{n+1/2}=1$ (with $t\neq 1$) is
 given by $\, \kappa_{2n+1}\cdot (x-x_0)^{(n+1)^2\, -1/2}$. 

\section{The direct sum structure}
\label{direct}
Perhaps the most striking feature of the linear differential operators for
${\tilde \chi}^{(3)}_d(t)$ and ${\tilde \chi}^{(4)}_d(t)$
exhibited in section (\ref{diffequ}) is their decomposition into a {\em direct
sum}. Such a decomposition has previously been seen for the full
susceptibility where the linear differential operator for ${\tilde \chi}^{(3)}$ is
the direct sum~\cite{jm1} of the  differential operator for
${\tilde \chi}^{(1)}$ and  a second linear differential operator. Similarly,  the linear 
operator for ${\tilde \chi}^{(4)}$ 
is the direct sum~\cite{jm3}
of the linear differential operator for ${\tilde \chi}^{(2)}$ and a
 second  linear differential operator.
In the papers~\cite{jm1}-\cite{jm3} the question is posed of how
general is the phenomenon of the direct sum decomposition.

In our previous paper~\cite{bm2} on the form factors $f^{(n)}(N,t)$ of
the diagonal Ising correlations we found that the linear differential operators of
all form factors for
$\, n\, \leq \,9$ have a direct sum decomposition, and that this decomposition is
surely valid for all values of $\, n$. 

An inspection of the direct sum decompositions reveals a great deal of
structure which is relevant to ${\tilde \chi}^{(n)}_d(t)$.
As an example consider the odd form factors $f^{(2n+1)}(N,t)$
relevant for $T\, >\, T_c$.
From the direct sum decomposition of the  linear 
differential operators~\cite{bm2}
for $\, f^{(2n+1)}(N, t)$, we find the following form:
\begin{eqnarray}
\label{sum2}
f^{(3)}(N,t)\, =\, \, \left({ N \over 2}+{1\over 6}\right) 
\cdot f^{(1)}(N,t)\,\, +g^{(3)}(N,t), 
\end{eqnarray}
\begin{eqnarray}
\label{sum3}
&&f^{(5)}(N,t)\, =\, \, \,\, 
{1\over 120}\cdot (15\, N^2\, +40\, N \, +9)\cdot f^{(1)}(N,t)\nonumber\\
&&~~~~~~~~~~~\quad \quad \quad \quad 
 +{1\over 2}( N +1)\cdot g^{(3)}(N,t)\,\, +g^{(5)}(N,t)
\end{eqnarray}
where $\, t^{N/2}\cdot  g^{(n)}(N,t)$ is a homogeneous polynomial of degree $n$
in the complete elliptic integrals $\, K$ and $\, E$, with coefficients which are
polynomials in $t$. If we then sum the decomposition (\ref{sum2}) of
$\, f^{(3)}(N,t)$, we find that
\begin{eqnarray}
\label{sum4}
{\tilde \chi}^{(3)}_d(t) \, = \,\, \,{\tilde \chi}^{(3)}_{d,1}(t)\, \,
+{\tilde \chi}^{(3)}_{d,2}(t)\, \, +{\tilde \chi}^{(3)}_{d,3}
\end{eqnarray}
where\footnote[5]{The integer $\, N$ should be understood as  $\,|N|$ in all the
summations below.}
\begin{eqnarray}
\label{sum41}
&&{\tilde \chi}^{(3)}_{d,1}(t)\, \,
=\,\, \,\, {1 \over 6}\sum_{N=-\infty}^{\infty}f^{(1)}(N,t)
\,\, =\,\, {1 \over 6}\, {\tilde \chi}^{(1)}_d, \\
\label{sum42}
&&{\tilde \chi}^{(3)}_{d,2}\, =\, \, \,
\sum_{N=-\infty}^{\infty}{ N \over  2}\cdot f^{(1)}(N,t)
\end{eqnarray}
where $ \, {\tilde \chi}^{(3)}_{d,2} $ is the solution of the linear
 differential operator $\, L^{(3)}_2$ (which is regular at
$ x=\, t^{1/2}=0$) and where
\begin{eqnarray}
\label{sumg3N}
{\tilde \chi}^{(3)}_{d,3} \, \,=\,\, \sum_{N=-\infty}^{\infty} \, g^{(3)}(N,t)
\, = \,\,\,\, b_1 \cdot {\tilde \chi}^{(1)}_d(t)\,\, +sol(L_3^{(3)})
\end{eqnarray}
with\footnote[8]{In fact, and since the full factorization of ${\cal L}_6^{(3)}$
is known, the projection of ${\tilde \chi}^{(1)}_d(t)$ in
${\tilde \chi}^{(3)}_d(t)$ can be computed and one finds $b_1=1/6$.}
   $\,b_1 \neq -1/6$. Thus we see that the direct sum decomposition of
the linear differential operators for ${\tilde \chi}^{(3)}_d(t)$ follows
immediately from the direct sum decomposition of the differential
operators for $\, f^{(3)}(N,t)$.

For $\, f^{(5)}(N, t)$, the sum (\ref{sum3}) can be written in terms
of $\, f^{(3)}(N, t)$ as
\begin{eqnarray}
&&f^{(5)}(N, t)\,  =\, \,\, 
-{\frac{1}{120}} \,\left(15\, N^2\,+1\right)\cdot  f^{(1)}(N, t) \nonumber \\
&& \qquad  \quad \quad \quad \, \quad
+ {\frac{1}{2}} \,(N+1)\cdot  f^{(3)}(N, t) \, + g^{(5)}(N, t) \nonumber
\end{eqnarray}
Summing on $N$, one obtains:
\begin{eqnarray}
\label{chi^{(5)}_d}
&&\chi^{(5)}_d \, = \, \, -{\frac{1}{120}} \,\sum_N f^{(1)}(N, t) \, 
- {\frac{1}{8}} \,\sum_N \, N^2 \cdot f^{(1)}(N, t)
\, + {\frac{1}{2}} \,\sum_N f^{(3)}(N, t) \nonumber \\
&&\quad \quad \quad +{\frac{1}{2}} \,\sum_N N \cdot f^{(3)}(N, t)\, 
\, +\sum_N g^{(5)}(N, t).
\end{eqnarray}
The first term, at the right-hand-side of (\ref{chi^{(5)}_d}), satisfies
the linear differential equation of order-one corresponding
to $\, \chi^{(1)}_d$, (i.e. $L^{(3)}_1$ in (\ref{ds3})).
The second term satisfies an order-one linear differential equation.
The third term is just $\chi^{(3)}_d$, up to a constant, a result
that we knew from the modulo prime method of section (\ref{Diffeqtilde5}).
We have found the differential operator corresponding to the
fourth term which is of order eight and has, in its direct sum, the linear
differential operator $\, L^{(3)}_2$ given in (\ref{ds3}).
As was the case for the sum in (\ref{sumg3N}), one might imagine
that the linear differential operator corresponding to the last term in
(\ref{chi^{(5)}_d}) will contain some of the differential operators
related to the previous terms.

We  consider now the ${\tilde \chi}^{(2n)}_d(t)$. 
In~\cite{bm2} we found that there is a direct sum decomposition for
the linear differential operator for $\, f^{(2n)}(N,t)$, just as there
was one for $f^{(2n+1)}(N,t)$.
From these direct sum decompositions, we find:
\begin{eqnarray}
\label{sum10}
&&f^{(2)}(N,t)\,=\,\, { N \over 2}\, +g^{(2)}(N,t), \\
\label{sum11}
&&f^{(4)}(N,t)\,=\,\,  { N  \cdot ( N +2)\over 8}\,
 +\left({ N \over 2}\, +{1\over 3}\right) \cdot g^{(2)}(N,t)
\,\, +g^{(4)}(N,t) \nonumber \\
\label{sum12}
&& \qquad \quad \,\,\, = \, \, { N \cdot (2-3 N)\over 24} + \,
\left({ N \over 2} +{1\over 3}\right) \cdot f^{(2)}(N,t) + g^{(4)}(N,t)
\end{eqnarray}

Similarly\footnote[2]{Note that the first terms in (\ref{sum10})
or (\ref{sum12})
do not lead to divergent sums. They are balanced by the sums on
the last terms, i.e., $g^{(2)}(N,t)$ and $g^{(4)}(N,t)$.},
the direct sum decomposition of the linear differential operators for
$\,f^{(2n)}(N,t)$ will lead to the
direct sum decomposition of the differential operators
for ${\tilde \chi}_d^{(2n)}(t)$ for all $n$.

\vskip 0.1cm

 We conclude that there will be a direct sum decomposition
for the linear differential operators for ${\tilde \chi}_d^{(n)}(t)$,
{\em for all} $n$, and that this direct sum decomposition 
is inherited from the direct sum decomposition for the linear differential
operators of the form factors of the Ising model.

\vskip 0.1cm

{\bf Remark:} It is worth noting that we have also performed
a large set of calculations (that will not be detailed here)
on more ``artificial'' toy susceptibilities like :
\begin{eqnarray}
\chi_{toy} \, = \, \,\,\,\,
 \sum_{j=1}^{\infty} \sum_{N=1}^{\infty} \, N^2 \cdot f^{(j)}(N,t).
\end{eqnarray}
Similarly, for the corresponding $ \, j$-particle contributions $\chi_{toy}^{(j)}$
we found for the first values of $j$ ($j\, = \, 1, \, 2, \, 3, \, 4$),
the corresponding Fuchsian linear differential operators.  
The singularities  of these Fuchsian linear differential equations
and of the corresponding $\, j$-fold integrals,
are totally and utterly similar to the one's of the diagonal susceptibility
analyzed in this paper. Again we have equations
totally similar to (\ref{sum4}), (\ref{sum41}), (\ref{sum42}).
This confirms, very clearly, that the direct sum decomposition for the 
$ \, j$-particle contributions $\chi_{toy}^{(j)}$ is 
straightforwardly inherited from the
direct sum decomposition for the form factors of the model.

\subsection{Resummations}
\label{resum}

There is one further feature of these decompositions which must be
mentioned. Namely, the sums of the form
\begin{eqnarray}
\sum_{N=-\infty}^{\infty} N ^p \cdot   f^{(2n+1)}(N, t)
\end{eqnarray}
will diverge at $t=1$ as $(1-t^{1/2})^{-p-1}$. Thus, for example, the
solution of $L^{(3)}_2$, given in (\ref{solL2}), diverges at $t^{1/2}=1$
as $(1-t^{1/2})^{-2}$. We see,  from (\ref{l33exp}), that the leading
singularity in the solution to $L^{(3)}_3$ will also diverge, at
$t^{1/2}=1$, as $(1-t^{1/2})^{-2}$. However, the full solution for
${\tilde \chi}_d^{(3)}(t)$ must diverge, at $t\rightarrow 1$, 
only as $(1-t^{1/2})^{-1}$, and, therefore, there must be cancellations
between terms in the direct sum. This phenomenon will happen for all
${\tilde \chi}_d^{(2n+1)}(t)$ where the cancellations become more
extensive as $n$ increases. 
For heuristic reasons let us consider $\chi^{(1)}_d$ versus $f^{(1)}(N)$.

The diagonal susceptibility of order one is defined as
\begin{eqnarray}
\label{chi1dsum}
\chi^{(1)}_d \, =\,  \sum_{N=-\infty}^{\infty} \, f^{(1)}(N, t)  \, =\,\,
\, f^{(1)}(0, t)\, \,\, +2\, \sum_{N=1}^{\infty} \, f^{(1)}(N, t)
\end{eqnarray}
with
\begin{eqnarray}
f^{(1)}(N, t)\,  = \,\,\, {\frac{(1/2)_N}{N!}}\cdot 
 t^{N/2} \cdot  {_2}F_1 \left( 1/2, 1/2+N; 1+N; t \right).
\end{eqnarray}

Writing in the sum over $N$, the hypergeometric function as a series, one obtains:
\begin{eqnarray}
\label{morechi1dsum}
\chi^{(1)}_d \, =\,\,\,
  K(t)\,
+ 2\, \sum_{k=0}^\infty \sum_{N=1}^\infty
{\frac{ (1/2)_k \,(1/2)_N \,(1/2+N)_k }{ (1+N)_k k! N!} } 
\cdot t^{k+N/2} 
\end{eqnarray}
with $(x)_k$ denoting the Pochhammer symbol, and $K(t)$ is as defined in (\ref{KE}).
One may shift $N$, separate the even from the odd $N$, and try to sum.
One may also generate the expansion, find the linear ODE and solve.
Remarkably the sum in (\ref{chi1dsum}) reduces to a simple expression
in terms of the complete elliptic integral of the first kind:
\begin{eqnarray}
\label{identity}
2\, \sum_{N=1}^{\infty} \, f^{(1)}(N, t) \, = \, \, \, 
\, -{\frac{\sqrt{t}}{t-1}} -{\frac{1}{t-1}} \, \,\, -K(t). 
\end{eqnarray}
This remarkable identity (\ref{identity}) explains why 
a sum like (\ref{chi1dsum}), where each term is polynomial expression of 
the complete elliptic integral of the first kind and of the second kind,
succeeds to reduce to a simple rational expression in $t^{1/2}$:
\begin{eqnarray}
\label{cancel}
\chi^{(1)}_d \,=\, \,\,  K\left( t \right) \, -{\frac{\sqrt{t}}{t-1}}
\,\, -{\frac{1}{t-1}} \,\, -K(t) \,\,=\, \,\, {\frac {1}{1-\sqrt{t}}}.
\end{eqnarray}

\section{The singularities in the diagonal susceptibility}
\label{behavior}

Thus far we have discussed the $\, n^{th}$ particle form factor contribution of
${\tilde \chi}_d^{(n)}(t)$. It remains to use this information to study the
diagonal susceptibility $\chi_d(t)$ itself, and,  for this,  
we need to consider
several problems which also occur for the computation of the full
susceptibility $\, \chi(s)$. 

We computed the susceptibilities $\chi_{d\pm}$ by summing the form
factor expansions (\ref{cdm}) and (\ref{cdp}) of the diagonal 
correlation functions $\, C(N,N)$ over all integer values
of $\, N$.
In doing this we have interchanged
the sum over position $\, N$ with the sum over form factors $n$. In field
theory language we have interchanged the high energy limit with the
sum over $n$ particle intermediate states.  This interchange is
universally done in both statistical mechanics and in field theory, but
should, in principle, be justified. 

Let us assume that this interchange of the sum over $N$ and $n$ can
be made. Then the behavior of $\chi_d$, as $\,T \,\rightarrow\,  T_c$, can be
studied from the behavior of ${\tilde \chi}^{(n)}(t)$ as $t\rightarrow
1$ if we make the additional assumption that the limit
$T\rightarrow T_c$ can also be interchanged with the sum over $n$.  
We then may  use (\ref{unit13a}) in (\ref{defchidm})  
to find, as $\, T \, \rightarrow \, T_c-$, that
\begin{eqnarray}
\label{tc1}   
kT \cdot \chi_{d-}\,  \,  \sim\, \,  \,  
(1-t)^{-3/4} \cdot \sum_{n=1}^{\infty}\, I_d^{(2n)}
\end{eqnarray}
and, similarly, by using (\ref{unit14a}) in (\ref{defchidp})  
we find,  as $T\rightarrow T_c+$, that:
\begin{eqnarray}
\label{tc2}
kT \cdot \chi_{d+} \, \,   \sim \, \,   \,  
(1-t)^{-3/4} \cdot \sum_{n=0}^{\infty}\, I_d^{(2n+1)}.
\end{eqnarray}
The sums in (\ref{tc1}) and (\ref{tc2}) must be shown to converge if
these estimates of the critical behavior are to be correct. For 
the full susceptibility, similar convergence has been recently demonstrated by
Bailey, Borwein and Crandall~\cite{crandall}.

The single pole divergence, which occurs in ${\tilde \chi}^{(n)}_d(t)$
for each $n$, is the analogue, for the diagonal susceptibility, of the
double pole divergence $(1-s)^2$ in the ${\tilde\chi}^{(n)}(s)$
of the full susceptibility. In both cases this divergence occurs from
terms in the integrand of ${\tilde \chi}^{(n)}_d(t)$ (or
 ${\tilde \chi}^{(n)}(s)$) which are not present in the corresponding integrals
for the form factor representation of the correlation function. These
divergences may be said to come from long distance effects, and are
captured in the scaling theory of the correlation functions. 


There are further singularities in ${\tilde \chi}_d^{(n)}(t)$
which come from the pinch singularities~\cite{jm6} of the square
root branch points $[(1-x)(1-tx)]^{1/2}$ in the integrands which
are also present in the form factors $f^{(n)}(N, t)$.
In the form factors $\, f^{(n)}(N, t)$, singularities give divergent
terms $\ln^n(1-t)$ which, term by term, would give the dominant
contribution $\, (1-t)^{1/4}\cdot \sum_n \ln^n(1-t)\,S_n(1-t) $ to the
correlation function.
However, the overall factor $(1-t)^{1/4}$ is absent in the diagonal
correlation $\, C(N,N)$ and is thus cancelled by the infinite sum on $n$.
We thus regain the original expansion of $\, C(N,N)$ as an $N\times N$
determinant whose singularities, at $t=1$, are of the form:
\begin{eqnarray}
\label{sing1}
(1-t)^{N^2}\cdot \ln^N(1-t).
\end{eqnarray}
These singularities in $\,C(N,N)$, which come from the
 summation of the form factors over $\,n$, may be said to be short distance singularities.

In ${\tilde \chi}^{(n)}_d(t)$, for $n=\,1,\,2$, the results of
section (\ref{evalu}) show that there are no logarithmic terms.
However, for $n=\, 3,\, 4$, logarithmic terms occur 
and we presume (but have not yet demonstrated)
for arbitrary $n>4$, that ${\tilde \chi}^{(n)}_d(t)$ will have 
logarithmic terms. 
However, just as was the case for $f^{(n)}(N, t)$,
these powers of log's must be summed over all $\, n$, and, again,
just as for $f^{(n)}(N, t)$, this sum must cancel the factor of
$\, (1-t)^{1/4}$. It will thus give terms of the form (\ref{sing1}) for
${\chi}_d(t)$ which are the counterpart of the similar terms in
the full susceptibility $\chi(s)$, and have been studied in detail 
by Orrick, Nickel, Guttmann and Perk~\cite{ongp}. 

It remains to discuss the singularities of the 
diagonal susceptibility on the unit circle $|t|=1$. 
These singularities not only have the property that they become
dense on the unit circle as $n\rightarrow \infty$, but 
since they are roots of unity, they accumulate according to
 a uniform distribution, in contrast with the
 density of the ``nickellian'' singularities 
(see (3.24) in~\cite{ongp} and (1) in~\cite{jm6}), 
which coincides
with the density of zeroes of the partition function without magnetic field
given in~\cite{LuWu}. 
Another qualitative difference
 between the diagonal susceptibility, and the full susceptibility,
is that these unit circle singularities for the diagonal susceptibility 
are logarithmic for $\, T\, <\,T_c$, and are square root type for $T\,>\,T_c$,
whereas, for the full susceptibility the singularities for $T\,>\,T_c$ are
logarithmic, and the singularities, for $T\, < \, T_c$, are of square root
type. In all cases the amplitudes of the singularities depend strongly
on $n$ which would seem to prevent any possible cancellation between
singularities in sets such as ${\tilde \chi}_d^{(2mn)}(t)$, with $n$
fixed and $m=\, 1,\, 2,\, \cdots$, which have the locations of some of the
singularities at coinciding positions. Therefore the arguments used by
Nickel \cite{nickel1}-\cite{nickel2} to conjecture a natural boundary 
in the full Ising
susceptibility will similarly suggest that there is a natural boundary  
in the diagonal Ising susceptibility as well. As is the case with the
full susceptibility a more rigorous argument would be most desirable.

\vskip .5cm 

\textbf{Acknowledgments:}
 We thank J-A. Weil and P. Flajolet for illuminating
 comments. We thank  A. J. Guttmann, I. Jensen,
 B. Nickel and R. Shrock for a large exchange of comments and ideas. 
 One of us (BM) has been partially 
supported by NSF grant DMR-0302758 and he  
thanks  the Depart. of Maths and Stats 
of the University of Melbourne where part of this work was 
performed. We acknowledge the support of a PICS/CNRS grant.
One of us (JMM) thanks  the MASCOS (Melbourne) where part of this work was 
performed.

\vskip .5cm 

\appendix

\section{Linear differential operators }
\label{B}

Recalling the factorization relation (\ref{russian})
 of the linear differential operator 
$\, {\cal L}_n^{(5)}$ corresponding to 
 ${\tilde \chi}_d^{(5)}(x)$ into a linear differential  operator
$\, {\cal L}_{n-6}^{(5)}$ and the  linear differential 
 operator corresponding to   ${\tilde \chi}_d^{(3)}(x)$:
\begin{eqnarray}
\label{russian2}
{\cal L}_n^{(5)}\, \, = \, \, \,\,  
{\cal L}_{n-6}^{(5)} \cdot {\cal L}_6^{(3)},
\end{eqnarray}
the smallest order for the linear differential operator $\, {\cal L}_{n-6}^{(5)}$ 
is thirteen, but this yields a large set of
 apparent singularities.
 For an order twenty, the linear differential  operator $\, {\cal L}_{20}^{(5)}$
 has no apparent singularities,
and requires less terms, in the series expansion, to be guessed.
Let us sketch this order twenty linear
 differential  operator $\, {\cal L}_{20}^{(5)}$
modulo the prime $32003$:
\begin{eqnarray}
\label{20}
&&{\cal L}_{20}^{(5)}\, = \, \, \,\, \,  
\sum_{i=0}^{i=20} \, Q_{i} \cdot Dx^{i}, \,
 \qquad \qquad \quad \hbox{with }     \\
&&Q_{20} \, = \,\, \left( 1+x \right)^{6} 
\left( {x}^{2}-1 \right)^{3} \left( {x}^{3}-1 \right)^{2}\,
 \left( {x}^{5}-1 \right)\, {x}^{10} \\
&& \qquad \, = \, \, (1+x)^{9} \, (x-1)^{6} \,
\, (1+x+{x}^{2})^{2} \, \,
 (1+x+{x}^{2}+{x}^{3}+{x}^{4}) \cdot {x}^{10} \nonumber \\
&&Q_{m} \, = \,\, (1+x)^{\alpha(9, \, m)} \cdot (x-1)^{\alpha(6, \, m)} \cdot
\, (1+x+{x}^{2})^{\alpha(2, \, m)} 
\cdot {x}^{\alpha(10, \, m)} \cdot q_{m}  \nonumber \\
&&  \hbox{where} \qquad \quad  m \, = \,\,  0,\, 1, \, 2, \,  \cdots, \, 19 
\qquad \quad \quad  \hbox{and}   \nonumber \\
&& \qquad \quad \alpha(N, \, m) \, = \, \, sup(0, \, N-20+m) \nonumber
\end{eqnarray}
and where the polynomials\footnote[2]{The polynomials  $ \,q_{m}$ 
are modulo the prime  $32003$.} $\,q_{m}$ read respectively :
\begin{eqnarray}
\label{20qm}
&&q_{19} \, = \, \, 29006 \cdot (x^2+1466\,x \, +22107) \cdot 
 P_{19}^{(17)} \cdot  R_{19}^{(17)} \cdot 
 P_{19}^{(13)} \cdot  P_{19}^{(6)},     \nonumber \\
&&q_{18} \, = \, \, 23383 \cdot (x+13780) \cdot (x+13647) \cdot  P_{18}^{(34)}
 \cdot  P_{18}^{(10)} \cdot  P_{18}^{(8)} \cdot  P_{18}^{(5)},  \nonumber \\
&&q_{17} \, = \, \, 23894 \cdot (x^2+13626\,x \, +7861) \cdot  P_{17}^{(59)},
\nonumber \\
&&q_{16} \, = \, \, 31481 \cdot (x+31392)\cdot  P_{16}^{(19)} 
\cdot  P_{16}^{(18)} \cdot  P_{16}^{(14)} \cdot  P_{16}^{(11)}, \nonumber \\
&&q_{15} \, = \, \, 13122 \cdot (x+2555) \cdot  P_{15}^{(40)} 
 \cdot  P_{15}^{(21)} \cdot  P_{15}^{(3)},  \nonumber \\
&&q_{14} \, = \, \, 1689 \cdot 
 P_{14}^{(57)}  \cdot  P_{14}^{(10)},   \nonumber \\
&&q_{13} \, = \, \, 851 \cdot (x+6919) \cdot  P_{13}^{(67)},   \nonumber \\
&&q_{12} \, = \, \, 542 \cdot 
 P_{12}^{(25)} \cdot  P_{12}^{(21)} \cdot  P_{12}^{(11)} 
\cdot  P_{12}^{(5)} \cdot  P_{12}^{(4)} \cdot  P_{12}^{(3)},
\nonumber \\
&&q_{11} \, = \, \, 26141 \cdot (x+1) \cdot P_{11}^{(58)} 
 \cdot P_{11}^{(5)} \cdot P_{11}^{(4)} \cdot P_{11}^{(2)},  \nonumber \\
&&q_{10} \, = \, \, 31757 \cdot x \cdot (x+14054) \cdot 
  P_{10}^{(55)} \cdot   P_{10}^{(11)} \cdot   P_{10}^{(2)},  \nonumber \\
&&q_{9} \, = \, \, 31477 \cdot  P_{9}^{(45)} \cdot  P_{9}^{(15)}
 \cdot  P_{9}^{(5)}\cdot  P_{9}^{(2)} \cdot  R_{9}^{(2)},  \nonumber \\
&&q_{8} \, = \, \, 28150 \cdot 
 P_{8}^{(62)} \cdot  P_{8}^{(4)} \cdot  P_{8}^{(2)}, \nonumber \\
&&q_{7} \, = \, \, 2111 \cdot 
(x+ 16608) \cdot  P_{7}^{(64)} \cdot  P_{7}^{(2)},  \nonumber \\
&&q_{6} \, = \, \, 21300 \cdot (x+ 15054)\cdot (x+ 20971) \cdot  P_{6}^{(37)}
\cdot  P_{6}^{(23)} \cdot  P_{6}^{(2)}\cdot  R_{6}^{(2)}, \nonumber \\
&&q_{5} \, = \, \,8699  \cdot (x+ 1134) \cdot   P_{5}^{(57)} 
\cdot   P_{5}^{(4)} \cdot  P_{5}^{(3)}, \nonumber \\
&&q_{4} \, = \, \, 7621  \cdot (x+ 27997) \cdot  P_{4}^{(7)} 
\cdot  P_{4}^{(42)}\cdot  P_{4}^{(8)}\cdot  P_{4}^{(6)}, \nonumber \\
&&q_{3} \, = \, \, 18283 \cdot 
(x+ 26460) \cdot  P_{3}^{(58)} \cdot  P_{3}^{(4)},   \nonumber \\
&&q_{2} \, = \, \, 2235 \cdot (x+ 8688) \cdot (x+ 20285) \nonumber \\
&& \quad \quad \times (x+ 24023) \cdot  P_{2}^{(34)} \cdot  P_{2}^{(12)}
\cdot  P_{2}^{(8)} \cdot  P_{2}^{(3)}  \cdot  P_{2}^{(2)},   \nonumber \\
&&q_{1} \, = \, \, 2139   \cdot (x+ 19284) \cdot (x+ 19339) 
\cdot  P_{1}^{(55)} \cdot  P_{1}^{(4)},   \nonumber \\
&&q_{0} \, = \, \, 23255 \cdot (x+ 30075)\cdot (x+ 13139) 
\cdot  P_{0}^{(41)} \cdot  P_{0}^{(15)} \cdot  P_{0}^{(2)}, \nonumber 
\end{eqnarray}
where the polynomials $\, P_{m}^{(N)}$
 or $\, R_{m}^{(N)}$  are polynomials of degree $\, N$ :
\begin{eqnarray}
P_{m}^{(N)} \, = \, \, \, \, x^N \, + \, \cdots 
\end{eqnarray}

\section{Singular behavior of  ${\tilde \chi}_d^{(3)}$
 and  ${\tilde \chi}_d^{(4)}$ }
\label{C}

Using the matrix connection method \cite{jm4}, we have obtained the
values of the amplitudes at the singularities,
lying on the unit circle $\vert t \vert =1$, of
${\tilde \chi}_d^{(3)}$ and ${\tilde \chi}_d^{(4)}$.

The linear differential equation for $\, \chi_d^{(3)}$ has the singularities
$x=\,0,\, 1,\, -1,\, \infty$ and the
 roots of $\, 1\, +x\, +x^2$ ($t=\,x^2$).
The singular behavior of $\,\chi_d^{(3)}$
at the singular
point $x_s$, 
 denoted by
$\chi_d^{(3)}\left( {\rm singular}, x_s \right)$ read (in the local variable $\,u=\,x-x_s$):
\begin{eqnarray}
\tilde{\chi}_d^{(3)}\left( {\rm singular}, 1 \right) \, =\,\,\,
 \left( {\frac{1}{3}}
\,+ {\frac{3}{4\pi}}\, \,+{\frac{a}{6}} \right) \cdot  {{1} \over {u}}\,\,
\,\, + {\frac{1}{8\, \pi}}\cdot  \ln(u) 
\end{eqnarray}
with $\, a =\, 0.469629259 \, \cdots$

\begin{eqnarray}
\tilde{\chi}_d^{(3)}\left( {\rm singular}, -1 \right)\,\, =\,\,\,
  {\frac{1}{4\pi^2}} \,\ln(u)^2\,
\,+ \left({\frac{1}{4\pi}}\,\, -{\frac{2\,\ln(2)\,-1}{2\, \pi^2}} \right)
 \cdot  \ln(u), \nonumber
\end{eqnarray}
\begin{eqnarray}
\tilde{\chi}_d^{(3)}\left( {\rm singular}, x_0 \right)\, =\,\,\,
  -{\frac{1}{6}} \,(1-i)\, (1+i\,\sqrt{3} )\cdot  b \cdot  u^{7/2} \nonumber
\end{eqnarray}
where $\, x_0=\, -1/2\,+i\,\sqrt{3}/2$ and with $\,b=\,0.203122784 \, \cdots$ 

The linear differential equation for $\tilde{\chi}_d^{(4)}$ has the singularities
$t=0, 1, -1, \infty$.
Denoting by $u=t-t_s$ the local expansion variable, where $t_s$ is the
singularity, the singular behavior for $\tilde{\chi}_d^{(4)}$ read
\begin{eqnarray}
&&\tilde{\chi}_d^{(4)}\left( {\rm singular}, 1 \right)\,  =\, \, 
 \left( {\frac{1}{\pi^2}}\,\, -{\frac{1}{4}}\,\,
 -{\frac{c}{8}} \right)  \cdot  {{1} \over {u}} 
\nonumber \\
&&\qquad \qquad \qquad  - {\frac{1}{8\,\pi^2}}\cdot  \ln(u)^2 \, \,
+{\frac{8\,\ln(2)\,-7\,-i\,2\,\pi}{8\, \pi^2}} \cdot \ln(u) \nonumber
\end{eqnarray}
with $\, c = \, -1.120950429 \, \cdots$
and:
\begin{eqnarray}
\tilde{\chi}_d^{(4)}\left( {\rm singular}, -1 \right)
\,\, = \,\,\,\, {\frac{1}{13440\, \pi^2}}
\cdot  u^7\cdot  \ln(u). \nonumber
\end{eqnarray}

\end{document}